\documentclass[letterpaper,oneside,twocolumn]{article}

\usepackage[hmargin=0.75in,vmargin=0.8in]{geometry}
\usepackage[utf8]{inputenc}
\usepackage{graphicx}
\usepackage{amsmath,mathtools}
\usepackage{amssymb}
\usepackage{amscd}
\usepackage{bm}
\usepackage{upgreek}
\usepackage{url}
\usepackage{hyperref}
\usepackage[super,sort&compress,comma]{natbib}
\usepackage[font=small,labelfont=bf,labelsep=period]{caption}
\usepackage[version=3]{mhchem}
\usepackage{booktabs}
\usepackage{multirow}
\usepackage{threeparttable}
\usepackage{sectsty}
\usepackage{setspace}
\usepackage{authblk}
\usepackage{fancyhdr}
\usepackage{enumitem}

\usepackage[T1]{fontenc}
\usepackage{textcomp}
\usepackage[euler]{textgreek}

\makeatletter
  \def\@seccntformat#1{\csname the#1\endcsname.\ \ }
\makeatother

\sectionfont{\large}
\subsectionfont{\normalsize}
\subsubsectionfont{\normalsize}

\makeatletter
 \def\@biblabel#1{#1.}
\makeatother
\DeclarePairedDelimiter{\nbr}{(}{)}
\DeclarePairedDelimiter{\cbr}{\{}{\}}
\DeclarePairedDelimiter{\sbr}{[}{]}
\DeclarePairedDelimiter{\abs}{\lvert}{\rvert}
\DeclarePairedDelimiter{\ensav}{\langle}{\rangle}

\renewcommand*{\vec}[1]{\ensuremath{\bm{#1}}}

\newcommand*{\dd}{\mathrm{d}}
\newcommand*{\rsub}[1]{\ensuremath{_{\mathrm{#1}}}}

\newcommand*{\pka}{\ensuremath{\mathrm{p}K_{\mathrm{a}}}}
\newcommand*{\degc}{\ensuremath{^\circ\mathrm{C}}}

\newcommand*{\inm}{\ensuremath{\mathrm{nm}^{-1}}}

\newcommand*{\gmola}{\ensuremath{\mathrm{g\;mol}^{-1}\:\text{\AA}^{-3}}}
\newcommand*{\e}{\ensuremath{\varepsilon}}
\newcommand*{\M}{\textsc{m}}
\newcommand*{\mM}{m\textsc{m}}

\makeatletter
\def\@maketitle{%
  \newpage
  \begingroup
  \let \footnote \thanks
  \hrule height \z@
    {\LARGE \bfseries \@title \par}%
    \vskip 3mm
    {\large
      \@author
    }%
  \par\endgroup
  \vskip 3mm 
  \vspace*{0mm}
}
\makeatother

\title{Weak self-interactions of globular proteins studied by small-angle X-ray
       scattering and structure-based modeling}

\author[1]{Shuji Kaieda}
\author[2]{Mikael Lund}
\author[3]{Tom\'as S.\ Plivelic}
\author[1]{Bertil Halle}

\affil[1]{Department of Biophysical Chemistry, Lund University, 
          Lund, Sweden}
\affil[2]{Department of Theoretical Chemistry, Lund University, 
          Lund, Sweden}
\affil[3]{MAX IV Laboratory, Lund University, 
          Lund, Sweden}
\date{}

\begin{document}

%
%
\twocolumn[
\begin{@twocolumnfalse}
\maketitle\thispagestyle{fancy}

%
%
\noindent 
We investigate protein--protein interactions in solution by small-angle X-ray 
scattering (SAXS) and theoretical modeling. 
The structure factor for solutions of bovine pancreatic trypsin inhibitor 
(BPTI), myoglobin (Mb), and intestinal fatty acid-binding protein (IFABP) is 
determined from SAXS measurements at multiple concentrations, from Monte Carlo 
simulations with a coarse-grained structure-based interaction model, and from 
analytic approximate solutions of two idealized colloidal interaction models 
without adjustable parameters. 
By combining these approaches, we find that the structure factor is essentially 
determined by hard-core and screened electrostatic interactions. 
Other soft short-ranged interactions (van~der~Waals and solvation-related) are 
either individually insignificant or tend to cancel out. 
The structure factor is also not significantly affected by charge fluctuations. 
For Mb and IFABP, with small net charge and relatively symmetric charge 
distribution, the structure factor is well described by a hard-sphere model. 
For BPTI, with larger net charge, screened electrostatic repulsion is also 
important, but the asymmetry of the charge distribution reduces the repulsion 
from that predicted by a charged hard-sphere model with the same net charge. 
Such charge asymmetry may also amplify the effect of shape asymmetry 
on the protein--protein potential of mean force. 
\vspace{5mm}
\end{@twocolumnfalse}
]

%
%
\section{\label{sec:intro}Introduction}

Protein--protein interactions govern the functional assembly of supramolecular 
structures\cite{Nooren2003,Keskin2008} as well as the dysfunctional aggregation 
of misfolded proteins.\cite{Chiti2006} 
Weak protein--protein interactions also determine the thermodynamics and phase 
behavior of concentrated protein solutions,\cite{Vekilov2010} of relevance for 
optimizing protein crystallization\cite{George1994} and for understanding how 
proteins behave in the crowded cytoplasm.\cite{Zhou2008} 
Fundamental progress in these areas requires a quantitative understanding of how 
proteins interact with themselves in solution. 
Specifically, we need to know the effective solvent-averaged protein--protein 
interaction energy or potential of mean force, $w\nbr{r}$. 

Much of the available information about protein--protein interactions in 
solution has come from scattering experiments via the osmotic second virial 
coefficient, $B_{22}$, and the structure factor, $S\nbr{q}$.\cite{Narayanan2003,
Zhang2007,Kim2008,Ianeselli2010,Cardinaux2011,Liu2011,Abramo2012,Heinen2012,
Zhang2012,Goldenberg2014} 
Whereas $B_{22}$ is an integral measure of the pair interaction, $S\nbr{q}$ is 
the Fourier transform of the isotropically averaged protein--protein pair 
correlation induced by the interactions.\cite{Hansen1986} 
Extraction of $w\nbr{r}$ from $S\nbr{q}$ is a nontrivial problem without a 
unique solution.\cite{Louis2011} 
Typically, a parameterized interaction model, $w\nbr{r;a,b,\ldots}$, is 
postulated and $S\nbr{q}$ is computed by molecular 
simulation\cite{Kim2008,Cardinaux2011,Abramo2012} or by an approximate integral 
equation theory.\cite{Narayanan2003,Zhang2007,Ianeselli2010,Liu2011,Heinen2012,
Zhang2012,Goldenberg2014}
The model parameters $a,b,\ldots$ are then optimized by comparing the computed $S\nbr{q}$ 
with that determined by small-angle X-ray (SAXS) or neutron (SANS) scattering. 

The interaction models used in this context may be classified as colloidal or 
structure-based. 
Colloidal interaction models are typically\cite{Narayanan2003,Zhang2007,
Ianeselli2010,Abramo2012,Heinen2012,Zhang2012} based on the 
Derjaguin--Landau--Verwey--Overbeek (DLVO) potential,\cite{Leckband2001} often 
complemented with phenomenological short-range contributions.\cite{Curtis1998} 
In the DLVO model, the protein is described as a uniformly surface-charged sphere 
embedded in a dielectric continuum. 
Such highly idealized models have the virtue of simplicity but cannot do full 
justice to protein--protein interactions.\cite{Neal1998,Piazza1999,Carlsson2001,
Grant2001,Allahyarov2002,Dahirel2007}
At the short and intermediate protein--protein separations, the irregular shape 
and the discrete and asymmetric charge distribution of real proteins cannot be 
ignored. 
Structure-based interaction models explicitly incorporate such structural 
features, either at atomic resolution or at a coarse-grained level. 
For computational expediency, the solvent is treated as a dielectric continuum; 
solvation-related interaction terms of a phenomenological nature are therefore 
sometimes included in the model. 
While this approach has been used extensively to compute 
$B_{22}$,\cite{Elcock2001,Lund2003,McGuffee2006,Persson2009,Mereghetti2010,
Stark2013,Quang2014} relatively few studies have reported $S\nbr{q}$ 
calculations with structure-based interaction 
models.\cite{McGuffee2006,Mereghetti2010} 
     
Here we report the structure factor $S\nbr{q}$, determined by SAXS, for aqueous 
solutions of three globular proteins: bovine pancreatic trypsin inhibitor 
(BPTI), equine skeletal muscle myoglobin (Mb), and rat intestinal fatty 
acid-binding protein (IFABP). 
To extract information about the protein--protein interactions, we use 
Metropolis Monte Carlo (MC) simulations to compute $S\nbr{q}$ for these 
solutions based on a coarse-grained structure-based (CGSB) interaction model 
with the individual amino acid residues as interaction sites.\cite{Lund2003} 
This implicit solvent model incorporates excluded volume, van~der~Waals (vdW) 
attraction, and screened Coulomb interactions, and the charges of the ionizable 
residues are allowed to fluctuate. 
To gain further insight, we compare the experimental and CGSB $S\nbr{q}$ with 
the (analytic) structure factors for two colloidal interaction models: the 
hard-sphere fluid in the Percus--Yevick (PY) 
approximation\cite{Wertheim1964,Nagele2004} and the hard-sphere Yukawa (HSY) 
fluid in the modified penetrating-background corrected rescaled mean spherical 
approximation (MPB-RMSA).\cite{Heinen2011,Heinen2011b} 

With only excluded volume and screened Coulomb interactions (no vdW attraction 
or other soft short-range interactions) and without any adjustable parameters, 
the CGSB model reproduces the experimental $S\nbr{q}$ nearly quantitatively for 
all three proteins within the $q$ range 0.5--3.0~$\inm$ accessed by the MC 
simulations. 
For Mb and IFABP, which were examined near isoelectric pH, the hard-sphere model 
predicts essentially the same $S\nbr{q}$ as does the CGSB model in this $q$ 
range. 
For the more highly charged BPTI, neither the hard-sphere model nor the charged 
hard-sphere model can reproduce the experimental $S\nbr{q}$. 
The implications of these findings are discussed.

%
%
\section{\label{sec:method}Materials and Methods}

%
%
\subsection{\label{subsec:saxs_exp}SAXS Experiments}

{\small
Protein solutions for SAXS measurements were prepared by dissolving lyophilized 
BPTI, Mb, or IFABP, purified and desalted as described,\cite{Kaieda2014} 
in MilliQ water. 
After adjusting pH by adding HCl or NaOH, the solutions were centrifuged at 
13\,000~rpm for 3~min to remove any insoluble protein. 
No buffers were used and the only electrolyte present is the counterions and a 
small amount of added salt (from pH adjustment) in the case of Mb. 
Relevant characteristics of the investigated protein solutions are summarized in 
Table~\ref{tab:sample}. 
}

\begin{table}[t!]
  \centering
  \footnotesize
  \begin{threeparttable}
    \caption{\label{tab:sample}\textbf{Characteristics of SAXS Samples}}
    \begin{tabular}{lcccccc}
      \toprule
      \multirow{2}{*}{protein} & $w\rsub{P}$ & $C\rsub{P}$ 
        & \multirow{2}{*}{$\phi\rsub{P}$ (\%)\tnote{a}} & \multirow{2}{*}{pH} 
        & \multirow{2}{*}{$Z\rsub{P}$\tnote{b}} & $C\rsub{salt}$ \\
      & (mg/ml) & (\mM) & & & & (\mM) \\
      \midrule
      \multirow{5}{*}{BPTI}  & 1.99 & 0.305  & 0.143  & 4.0 & +7.4 & 0     \\[1ex]
                             & 9.75 & 1.50   & 0.702  & 4.0 & +7.4 & 0     \\[1ex]
                             & 39.9 & 6.12   & 2.87   & 4.1 & +7.2 & 0     \\[1ex]
                             & 101  & 15.5   & 7.27   & 4.1 & +7.2 & 0     \\
      \midrule
      \multirow{4}{*}{Mb}    & 1.32 & 0.0752 & 0.0979 & 6.8 & +3.2 & 0.188 \\[1ex]
                             & 8.43 & 0.480  & 0.625  & 6.8 & +3.2 & 1.20  \\[1ex]
                             & 29.0 & 1.65   & 2.15   & 6.8 & +3.2 & 4.13  \\
      \midrule
      \multirow{5}{*}{IFABP} & 7.77 & 0.501  & 0.567  & 7.0 & +0.2 & 0     \\[1ex]
                             & 15.5 & 1.00   & 1.13   & 7.0 & +0.2 & 0     \\[1ex]
                             & 31.0 & 2.00   & 2.26   & 7.0 & +0.2 & 0     \\[1ex]
                             & 62.0 & 4.00   & 4.53   & 7.0 & +0.2 & 0     \\
      \bottomrule
    \end{tabular}
    \footnotesize
    \begin{tablenotes}[flushleft]
      \item[a] The protein volume fraction was obtained as 
               $\phi\rsub{P} = n\rsub{P} V\rsub{P}$, with $n\rsub{P}$ being the 
               protein number density and $V\rsub{P}$ the protein (partial) 
               volume (see text). 
      \item[b] Net protein valency, calculated with experimental $\pka$ values 
               when available (Asp, Glu, Lys, Tyr, and N- and C-termini for 
               BPTI\cite{March1982} and His for Mb\cite{Kao2000}) and with 
               standard $\pka$ values in proteins otherwise (C-terminus, 2.5; Asp, 3.65; 
               Glu, 4.45; His, 6.5; N-terminus, 8.0; Tyr, 10.0; Lys, 10.6; 
               Arg, 12.5). 
    \end{tablenotes}
  \end{threeparttable}
 \normalsize
\end{table}

{\small
SAXS measurements were performed at the MAX-lab synchrotron beamline I911-4, 
equipped with a PILATUS 1M detector (Dectris).\cite{Labrador2013} 
The scattering vector $q$ range ($q = 4\pi/\lambda\:\sin\theta$, where 
$\lambda = 0.91$~\AA\ is the X-ray wavelength and $2\theta$ is the scattering 
angle) was calibrated with a silver behenate sample. 
All measurements were performed on samples in flow-through cells at 20$\degc$ 
with an exposure time of 1~min. 
The effect of radiation damage did not exceed the experimental noise. 
Reported scattering profiles $I\nbr{q}$ were obtained as the difference of the 
azimuthally averaged 2D SAXS images from protein solution and solvent (MilliQ 
water).
}

%
%
\subsection{\label{subsec:saxs_anal}SAXS Data Analysis}

{\small
For a solution of $N\rsub{P}$ protein molecules of volume $V\rsub{P}$ contained 
in a volume $V$, the scattering intensity $I\nbr*{q}$ in the decoupling 
approximation, where the orientation of a protein molecule is taken to be 
independent of its position and the configuration of other protein molecules, 
can be factorized as\cite{Guinier1955,Kratky1982,Pedersen1997} 
\begin{equation}
  I\nbr*{q} = n\rsub{P}\nbr*{V\rsub{P} \Delta\rho}^2 P\nbr*{q} S\nbr*{q}\ ,
  \label{eq:iq}
\end{equation}
where $n\rsub{P} = N\rsub{P}/V$ is the protein number density, $\Delta\rho$ is 
the protein--solvent electron density difference (the scattering contrast), 
$P\nbr{q}$ is the form factor, and $S\nbr{q}$ is the structure factor. 
Because of the non-spherical protein shape, Eq.~\eqref{eq:iq} should involve an 
effective structure factor $\bar{S}\nbr{q}$, which, however, differs 
insignificantly from $S\nbr{q}$ under the conditions of the present study. 
The form factor represents the scattering from an isolated protein molecule, 
\begin{equation}
  P\nbr*{q} = \ensav*{\abs*{\frac{1}{V\rsub{P}} 
    \int_{V\rsub{P}} \dd\vec{r} \exp\nbr*{-i \vec{q}\cdot\vec{r}}}^2}\ ,
  \label{eq:pq}
\end{equation}
whereas the structure factor reflects intermolecular pair correlations, 
\begin{equation}
  S\nbr*{q} = \sum_{k=1}^{N\rsub{P}}  
    \ensav*{\exp\sbr*{-i \vec{q}\cdot\nbr*{\vec{r}_1-\vec{r}_k}}}\ .
  \label{eq:sq}
\end{equation}
In Eqs.\ \eqref{eq:pq} and~\eqref{eq:sq}, $\ensav{\ldots}$ signifies an 
equilibrium configurational average. 
}

{\small
According to Eq.~\eqref{eq:iq}, the structure factor, $S\nbr{q;n\rsub{P}}$, at a 
protein concentration $n\rsub{P}$ can be obtained by dividing the 
concentration-normalized intensity, $I\nbr{q;n\rsub{P}}/n\rsub{P}$, by the 
same quantity measured at a sufficiently low concentration, $n\rsub{P}^0$, that 
$S\nbr{q;n\rsub{P}^0} \equiv 1$. 
We shall refer to 
$I\nbr{q;n\rsub{P}^0}/n\rsub{P}^0 = \nbr{V\rsub{P} \Delta\rho}^2 P\nbr*{q}$ 
as the apparent form factor (AFF). 
As described in more detail elsewhere,\cite{Kaieda2014} the AFF for each protein 
was constructed by merging concentration-normalized SAXS profiles from two 
different protein concentrations (the highest and the lowest in 
Table~\ref{tab:sample}) and by smoothing the merged profile. 
The low $q$ part of the AFF, where the SAXS profile is sensitive to 
protein--protein correlations, originates from the  dilute solution with 
$S\nbr{q} \approx 1$, whereas the high $q$ part, which reflects intraprotein 
correlations, is derived from a concentrated solution with better 
signal-to-noise. 
}

%
%
\subsection{\label{subsec:mc}CGSB Interaction Model and MC Simulation}

\begin{figure}[t!]
  \centering
  \includegraphics[viewport=0 0 229 340]{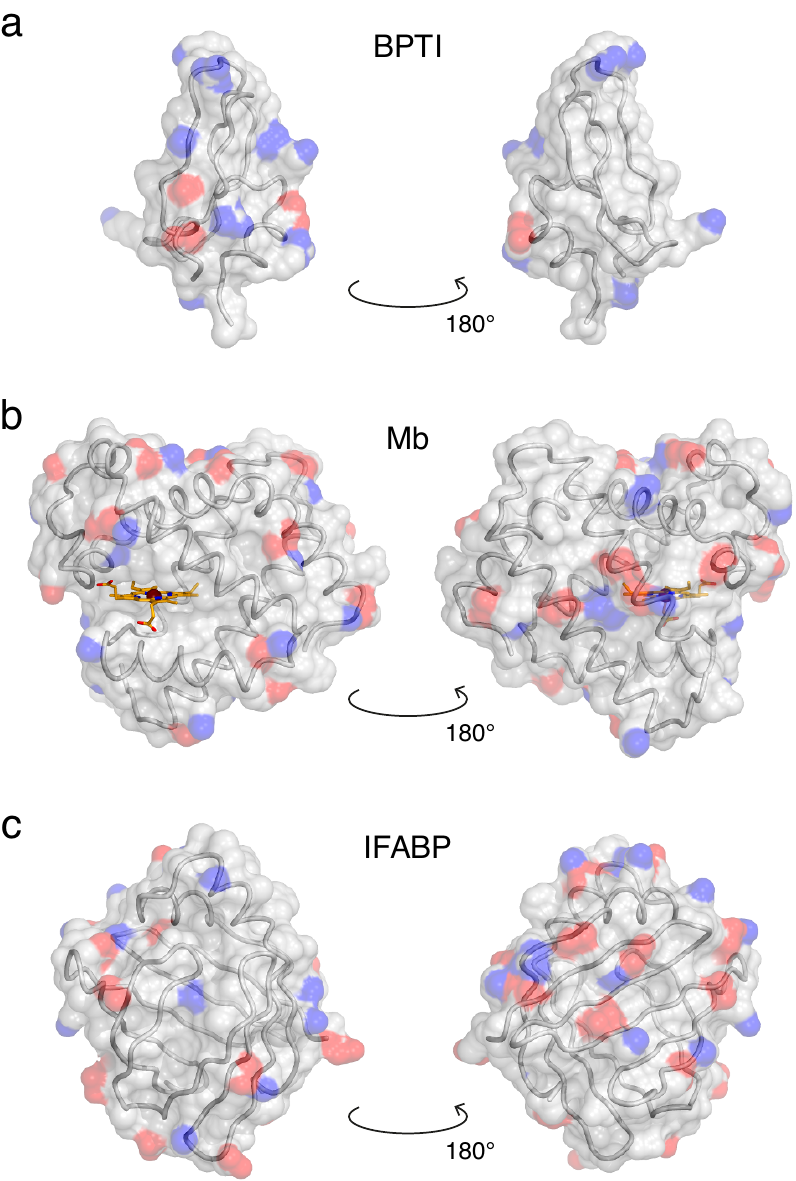}
  \caption{\label{fig:struct}Crystal structures of BPTI (\textbf{a}; PDB ID 
           1bpi\cite{Parkin1996}), Mb (\textbf{b}; 1wla\cite{Maurus1997}), and 
           IFABP (\textbf{c}; 1ifc\cite{Scapin1992}). 
           Backbone and surface representations are superimposed and the heme 
           group of Mb is shown in stick representation. 
           The protein surfaces are colored red or blue at the positions of Asp 
           and Glu O atoms and Lys and Arg N atoms, respectively. 
           The figure was prepared with CueMol (\url{http://www.cuemol.org}).}
\end{figure}

{\small
In the CGSB interaction model, each amino acid residue (plus the terminal amino 
and carboxyl groups) is represented by an isotropic interaction site, placed at 
the center-of-mass of the corresponding residue in the crystal structure of the 
real protein (Fig.~\ref{fig:struct}). 
(For simplicity, we shall refer to these interaction sites as residues.) 
The effective energy of interaction between residues $i$ and $j$, separated by 
a distance $r_{ij}$, is taken to be 
\begin{equation}
  u\nbr*{r_{ij}} 
    = k\rsub{B} T \lambda\rsub{B} \frac{z_i z_j}{r_{ij}} 
        e^{-\kappa r_{ij}}
      + 4 \e \sbr*{\nbr*{\frac{\sigma_{ij}}{r_{ij}}}^{\!\!12} 
                   \!\!- \nbr*{\frac{\sigma_{ij}}{r_{ij}}}^{\!\!6}} 
      + \delta_{ij}\nbr*{r\rsub{C}}\ .
  \label{eq:potential}
\end{equation}
The first term describes the electrostatic interaction in the Debye--H\"uckel 
approximation.
Here, $\lambda\rsub{B} = 0.71$~nm is the Bjerrum length for water at 20$\degc$, 
$\kappa = \nbr{4\pi \lambda\rsub{B} \abs{Z\rsub{P}} n\rsub{P}}^{1/2}$ is the 
inverse Debye screening length determined by the counterions (no added salt) of 
the protein with net charge valency $Z\rsub{P}$, and $z_i = 0$ or $\pm 1$ is the 
valency of residue $i$. 
The second term in Eq.~\eqref{eq:potential}, a Lennard-Jones (LJ) potential with 
well depth $\e$ and $\sigma_{ij} = \nbr{\sigma_i+\sigma_j}/2$, describes 
exchange repulsion and vdW attraction. 
The vdW diameter $\sigma_i$ was fixed by the residue molar mass, $M_i$, 
according to $\sigma_i = \sbr{6 M_i/\nbr{\pi \rho}}^{1/3}$ with 
$\rho = 1~\gmola$. 
(Varying the density  $\rho$ by $\pm 20\%$ has negligible effect on the 
structure factor.)
Finally, in the third term of Eq.~\eqref{eq:potential}, 
$\delta_{ij}\nbr{r\rsub{C}}$ shifts the pair potential to zero at a spherical 
cut-off distance $r\rsub{C}$ in the rang 0.1--5~$\kappa^{-1}$ (4.8--27.2~nm). 
Relevant characteristics of the simulated protein solutions are collected in 
Table~\ref{tab:mcdata}. 
}

\begin{table}[t!]
  \centering
  \footnotesize
  \begin{threeparttable}
    \caption{\label{tab:mcdata}\textbf{Characteristics of Simulated Solutions}}
    \begin{tabular}{llcccccc}
      \toprule
      \multirow{2}{*}{protein} & \multirow{2}{*}{PDB} 
        & \multirow{2}{*}{$N\rsub{res}$\tnote{a}} & \multirow{2}{*}{pH} 
        & $C\rsub{P}$ & \multirow{2}{*}{$Z\rsub{P}$\tnote{b}} & $\kappa^{-1}$ 
        & $r\rsub{C} \kappa$ \\
      & & & & (\mM) & & (nm) & \\
      \midrule
      \multirow{3}{*}{BPTI} & \multirow{3}{*}{1bpi\cite{Parkin1996}} 
        & \multirow{3}{*}{58} & \multirow{3}{*}{4.1} 
            & 1.50  & $+6.3$          & 4.37 & 5   \\[1ex]
      & & & & 6.12  & $+6.6$          & 2.16 & 5   \\[1ex]
      & & & & 15.5  & $+7.0$          & 1.36 & 5   \\
      \midrule
      \multirow{2}{*}{Mb} & \multirow{2}{*}{1wla\cite{Maurus1997}} 
        & \multirow{2}{*}{153} & \multirow{2}{*}{6.8} 
            & 0.480 & $+2.0$\tnote{c} & 13.6  & 2   \\[1ex]
      & & & & 1.65  & $+2.1$\tnote{c} & 7.29 & 2   \\
      \midrule
      \multirow{3}{*}{IFABP} & \multirow{3}{*}{1ifc\cite{Scapin1992}} 
        & \multirow{3}{*}{131} & \multirow{3}{*}{7.0} 
            & 1.00  & $-0.021$        & 96.3 & 0.1 \\[1ex]
      & & & & 2.00  & $-0.021$        & 68.1 & 0.1 \\[1ex]
      & & & & 4.00  & $-0.017$        & 48.1 & 0.1 \\
      \bottomrule
    \end{tabular}
    \footnotesize
    \begin{tablenotes}[flushleft]
      \item[a] Number of residues per protein. 
               The number of interaction sites is $N\rsub{res} + 2$. 
      \item[b] Average net protein valency determined from the simulation. 
      \item[c] The fixed valency of the heme group in Mb was set to $+1$. 
    \end{tablenotes}
  \end{threeparttable}
  \normalsize
\end{table}

\begin{figure}[t!]
  \centering
  \includegraphics[]{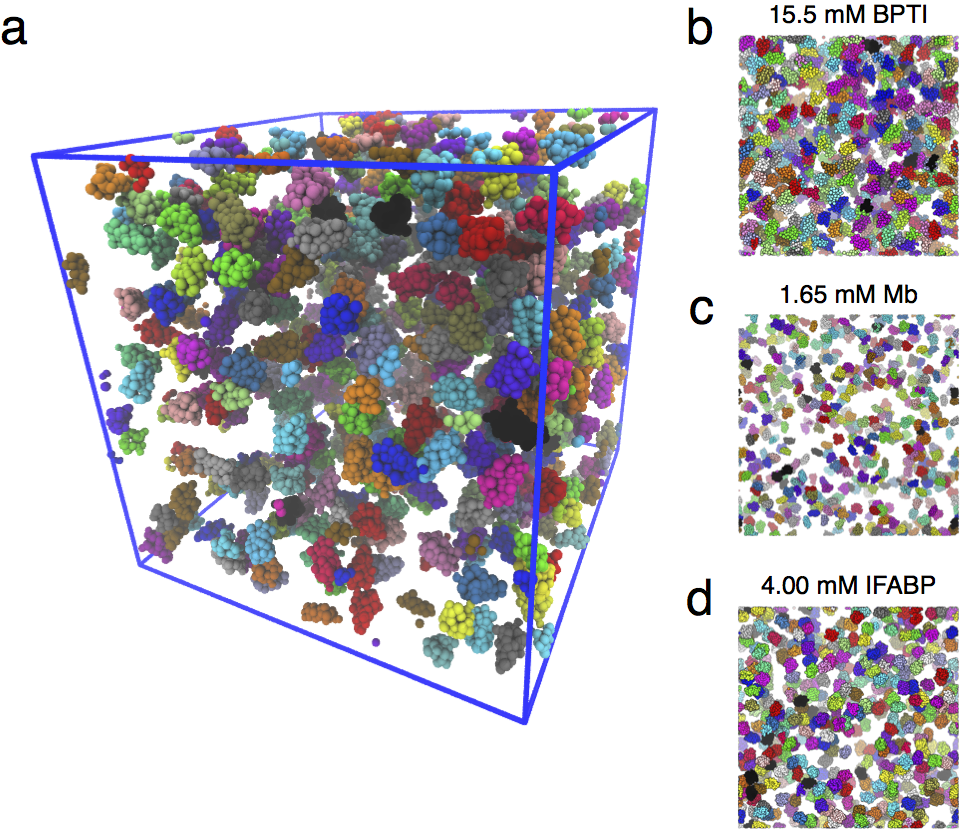}
  \caption{\label{fig:snapshot}Snapshots from MC simulations. 
           (\textbf{a}) 500 BPTI molecules (30\,000 interaction sites) in a 
           cubic cell at $C\rsub{P} = 15.5$~\mM. 
           (\textbf{b})--(\textbf{d}) Side-view of the most concentrated 
           solution simulated for each protein (Table~\ref{tab:mcdata}).}
\end{figure}

{\small
MC simulations were performed at 293~K in the $NVT$ ensemble with fluctuating 
protein charges (constant pH) using the Faunus framework.\cite{Stenqvist2013} 
The cubic simulation box, with periodic boundary conditions, contained 
$N\rsub{P} = 500$ rigid, coarse-grained protein molecules and the box volume was 
adjusted to match the experimental protein concentrations 
(Table~\ref{tab:mcdata} and Fig.~\ref{fig:snapshot}). 
Configurational space, that is, the position and orientation of each protein 
molecule and the protonation state of each ionizable group, was sampled by the 
conventional Metropolis algorithm\cite{Hastings1970} using the following energy 
function, 
\begin{equation}
  U = \sum_{i} \sum_{j>i} u\nbr*{r_{ij}} 
      + k\rsub{B} T \ln 10 \sum_{n} \alpha_ n
          \nbr*{\mathrm{pH}-\mathrm{p}K^\circ_{\mathrm{a},n}}\ . 
  \label{eq:hamiltonian}
\end{equation}
In the first term, $u\nbr{r_{ij}}$ is the pair potential from 
Eq.~\eqref{eq:potential} and the double sum runs over all pairs of residues (in 
the same or in different protein molecules). 
In the second term, which ensures that the fluctuating charges conform to a 
Boltzmann distribution,\cite{Sassi1992,Ullner1994} the sum runs over all 
ionizable residues and $\alpha_n = 1$ or 0 for residues in protonated and 
deprotonated forms, respectively. 
The intrinsic (in the absence of electrostatic interactions) 
$\mathrm{p}K^\circ_{\mathrm{a},n}$ was taken to be 3.8 (C-terminus), 4.0 (Asp), 
4.4 (Glu), 6.3 (His), 7.5 (N-terminus), 9.6 (Tyr), 10.4 (Lys), or 12.0 (Arg). 
Shifts in the apparent acid dissociation constant, $\mathrm{p}K_{\mathrm{a},n}$, 
due to intramolecular and intermolecular electrostatic interactions are 
explicitly accounted for by the first term in Eq.~\eqref{eq:potential}. 
Charge fluctuations give rise to a short-ranged attractive protein--protein 
interaction.\cite{Kirkwood1952,Lund2013} 
}

{\small
During the simulation, the rigid protein molecules were subjected to combined 
mass-center translations and rotations (25\,000 moves per protein molecule), 
while the protonation state of all ionizable residues were alternated between 
protonated and deprotonated forms (20\,000 moves per protein molecule). 
Each production MC run was preceded by a tenfold shorter equilibration run. 
From the MC-generated ensemble of equilibrium configurations, we computed the 
average net protein valency, $Z\rsub{P} = \ensav{\sum_n z_n}$ 
(Table~\ref{tab:mcdata}), and the isotropically averaged static structure factor, 
$S\nbr{q}$. 
The latter was computed from the Debye formula,\cite{Guinier1955,Kratky1982} 
\begin{equation}
  S\nbr*{q} = 1 
    + \frac{2}{N\rsub{P}} \ensav*{\sum_{i=1}^{N\rsub{P}-1} 
      \sum_{j=i+1}^{N\rsub{P}} \frac{\sin\nbr*{qR_{ij}}}{qR_{ij}}}\ ,
  \label{eq:debye}
\end{equation}
where the double sum runs over all unique protein mass-center separations, 
$R_{ij}$. 
The $q$ range of the calculated $S\nbr{q}$ is limited to $> 0.5~\inm$ due to 
the finite size of the simulation box.
}

%
%
\subsection{\label{subsec:colloid}Colloidal Interaction Models}

{\small
Two colloidal interaction models were examined, both of which describe the 
protein as a spherical particle. 
In both cases, we used analytic expressions for $S\nbr{q}$ obtained from 
approximate but accurate solutions of the Ornstein--Zernike integral 
equation.\cite{Hansen1986} 
For the hard-sphere fluid, where excluded volume is the only interaction, we 
used the PY approximation,\cite{Wertheim1964,Nagele2004} which is virtually 
exact for a hard-sphere fluid at the volume fractions of interest here. 
The HSY fluid includes, in addition to hard-core repulsion, a screened Coulomb 
(Yukawa) interaction between two uniformly charged spheres. 
For this model, we used the MPB-RMSA,\cite{Heinen2011,Heinen2011b} which yields 
$S\nbr{q}$ in excellent agreement with simulations (for this model) over the 
full parameter space.\cite{Heinen2011,Heinen2011b} 
For convenience, we reproduce the analytic $S\nbr{q}$ expressions for these two 
models in Supporting Information (Secs.\ \ref{sec:hs} and~\ref{sec:yukawa}).
}

{\small
As in the case of the CGSB model, we did not fit any of the parameters in the 
colloidal interaction models. 
The hard-sphere diameter, $\sigma\rsub{P}$, was set to 2.46, 3.46, and 3.30~nm 
for BPTI, Mb, and IFABP, respectively, which reproduce the actual protein 
volumes, $V\rsub{P}$, of 7.79, 21.7, and 18.8~nm$^3$, respectively, obtained 
from the molar mass and partial specific volume of these 
proteins.\cite{Filfil2004,DeMoll2007}
The protein volume fraction, $\phi\rsub{P}$, and net valency, $Z\rsub{P}$, were 
set to the values given in Tables \ref{tab:sample} and~\ref{tab:mcdata}, 
respectively. 
}

%
%
\section{\label{sec:result}Results and Discussion}

%
%
\subsection{\label{subsec:expsq}Structure Factor from SAXS}

\begin{figure}[t!]
  \centering
  \includegraphics[viewport=0 0 209 428]{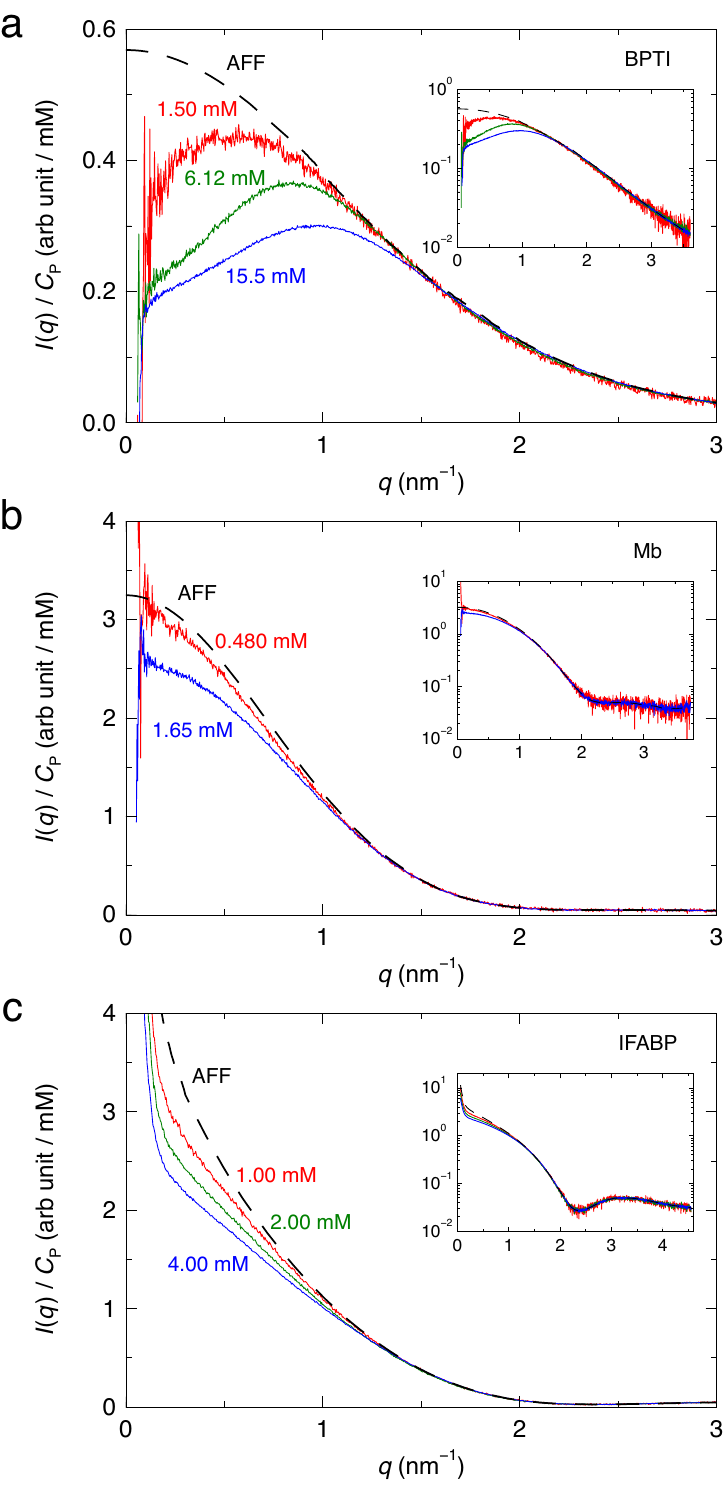}
  \caption{\label{fig:profile}Concentration-normalized SAXS profiles from 
           solutions of BPTI (\textbf{a}), Mb (\textbf{b}), and IFABP 
           (\textbf{c}) at different concentrations (solid curves). 
           Also shown is the AFF for each protein (dashed curve). 
           The insets show the same data in semilog format.}
\end{figure}

\begin{figure}[t!]
  \centering
  \includegraphics[viewport=0 0 207 428]{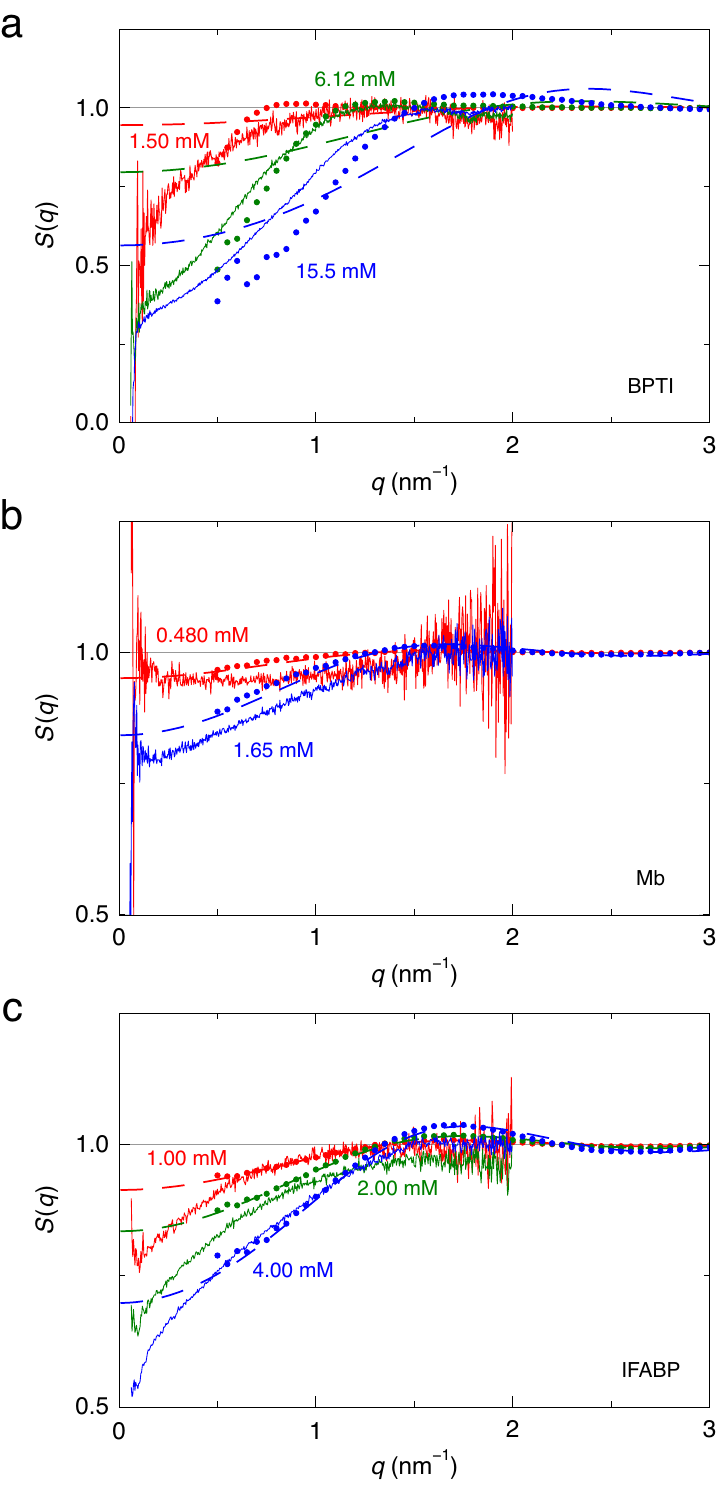}
  \caption{\label{fig:sqpy}Structure factor for BPTI (\textbf{a}), 
           Mb (\textbf{b}), and IFABP (\textbf{c}) solutions at several 
           concentrations, obtained from SAXS experiments (solid curves),
           from the CGSB model without vdW attraction (dots), and from
           the hard-sphere model (dashed curves). 
           The experimental $S\nbr{q}$ is only shown up to $q = 2~\inm$; 
           at higher $q$ the noise amplitude exceeds any deviation from 
           $S\nbr{q} = 1$.}
\end{figure}

Excess (protein solution minus water) scattering profiles, $I\nbr{q}$, were 
obtained from SAXS measurements on solutions of BPTI, Mb, and IFABP at several 
concentrations. 
In Fig.~\ref{fig:profile} we have divided $I\nbr{q}$ by the protein molar 
concentration, $C\rsub{P}$, to remove the trivial concentration dependence 
(see Eq.~\eqref{eq:iq}). 
As expected, $I\nbr{q}/C\rsub{P}$ is independent of $C\rsub{P}$ at high $q$, 
where intramolecular scattering dominates. 
At lower $q$ values, $I\nbr{q}/C\rsub{P}$ decreases with increasing $C\rsub{P}$, 
indicating predominantly repulsive protein--protein interactions. 
The structure factor, $S\nbr{q}$, in Fig.~\ref{fig:sqpy} was obtained, as 
described in Sec.~\ref{subsec:saxs_anal}, by dividing $I\nbr{q}/C\rsub{P}$ with 
the AFF, also shown in Fig.~\ref{fig:profile}. 

Under certain solution conditions (high pH, high salt concentration), BPTI 
exists in an equilibrium between monomeric and decameric 
forms.\cite{Hamiaux2000,Gottschalk2003} 
Since the pronounced minima at $q = 1.5$ and 2.9~$\inm$ in the decamer form 
factor\cite{Hamiaux2000,Kaieda2014} are not evident in our SAXS profiles 
(Fig.~\ref{fig:profile}a), we conclude that decamers are not present in our 
BPTI solutions. 
The large intensity increase at $q \lesssim 0.2~\inm$ seen in all IFABP profiles 
(Fig.~\ref{fig:profile}c) can be explained by a small fraction 
($\sim 10^{-5}$) of protein in large aggregates (effective diameter 
$\sim 10 \times \sigma\rsub{P}$). 
Rather than treating this structural heterogeneity explicitly, we incorporate 
the aggregate contribution in the AFF. 
To the extent that aggregation is concentration-dependent, this procedure may 
introduce artifacts in $S\nbr{q}$ at $q \lesssim 0.2~\inm$. 
Apart from this anomaly in the IFABP profiles, the AFFs for all three proteins 
agree well with the form factors computed with the \textsc{crysol} 
program\cite{Svergun1995} from the corresponding crystal structures 
(Fig.~\ref{fig:struct}).

%
%
\subsection{\label{subsec:mcsq}Structure Factor from CGSB Model}

Figure~\ref{fig:sqpy} also shows the structure factor predicted by the CGSB 
interaction model. 
This structure factor was computed from MC simulations at the experimental 
temperature, pH, and protein concentrations and with the structural model 
parameters determined by the protein crystal structures 
(Fig.~\ref{fig:struct}).
The only parameter that is not fixed by the protein structure is the LJ well 
depth $\e$ (see Eq.~\eqref{eq:potential}). 
Nominally, this parameter measures the strength of the average residue--residue 
vdW attraction across the aqueous solvent, but, in practice, it may also subsume 
short-range solvation-related interactions that are not explicitly accounted for 
in the CGSB model. For the CGSB calculations shown in Fig.~\ref{fig:sqpy}, we 
have set $\e = 0.005~k\rsub{B}T$, corresponding to a negligibly weak apparent 
vdW interaction. 
(We cannot set $\e = 0$ since this parameter also scales the steep repulsive 
term in Eq.~\eqref{eq:potential}, which is essentially determined by the vdW 
contact separations, $\sigma_{ij}$.) 

\begin{figure}[t]
  \centering
  \includegraphics[viewport=0 0 197 137]{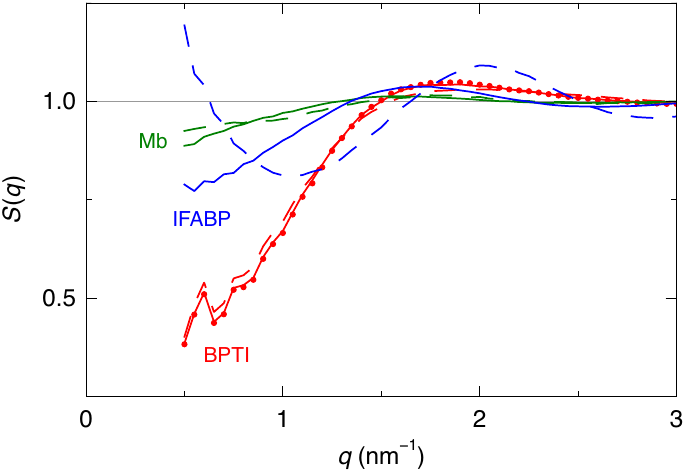}
  \caption{\label{fig:vdw}Structure factor predicted by the CGSB model for BPTI, 
           Mb, and IFABP at the highest concentrations in Table~\ref{tab:mcdata} 
           and with ($\e = 0.05~k\rsub{B}T$, dashed curves) or without 
           ($\e = 0.005~k\rsub{B}T$, solid curves) vdW attraction. 
           Also shown is $S\nbr{q}$ for BPTI from a simulation with fixed 
           charges and no vdW attraction (dots).}
\end{figure}

The qualitative, and in some cases semi-quantitative, agreement found, in the 
$q$ range ($> 0.5~\inm$) accessed by the MC simulations, between the structure 
factors predicted by the CGSB model with $\e = 0.005~k\rsub{B}T$ and measured by 
SAXS (Fig.~\ref{fig:sqpy}) indicates that the solution structure can be 
fairly well described by an interaction model that only incorporates excluded 
volume and screened inter-residue Coulomb interactions. 
In other words, the vdW attraction and other short-range soft interactions are 
either individually negligibly weak or tend to cancel out. 
A tenfold increase of the vdW attraction to $\e = 0.05~k\rsub{B}T$, as used in 
previous applications of the CGSB 
model,\cite{Lund2003,Persson2009,Persson2009b,Kurut2012} has little effect on 
$S\nbr{q}$ at $q > 0.5~\inm$ for the two proteins (BPTI and Mb) with significant 
net charge (Fig.~\ref{fig:vdw}). 
In contrast, a large effect is seen for IFABP (Fig.~\ref{fig:vdw}), likely 
because the electrostatic repulsion close to the isoelectric pH 
(Table~\ref{tab:mcdata}) is so weak that the protein molecules come into vdW 
contact more frequently.    

The MC simulations with the CGSB model were carried out at constant pH. 
The protonation state of ionizable residues therefore undergoes thermal 
fluctuations and responds to the local electrostatic potential produced by 
charged residues in the same protein molecule and in nearby protein molecules. 
However, even for BPTI, which was studied at a pH where charge fluctuations are 
large (close to the $\pka$ of carboxyl groups), the attractive electrostatic 
interaction produced by charge fluctuations\cite{Kirkwood1952,Lund2013} has 
negligible effect on the structure factor (Fig.~\ref{fig:vdw}). 
For Mb and IFABP, which were studied near neutral pH where charge fluctuations 
are less pronounced, the effect of charge fluctuations on $S\nbr{q}$ should be 
even smaller.

In the fluctuating-charge CGSB model, the protonation state of ionizable 
residues is affected by intramolecular and intermolecular electrostatic 
interactions. 
For all three proteins, the net protein charge, $Z\rsub{P}$, computed from this 
model (Table~\ref{tab:mcdata}) is within one unit from the $Z\rsub{P}$ value 
obtained with experimental $\pka$ values (Table~\ref{tab:sample}).
We find that $Z\rsub{P}$ depends weakly on protein concentration 
(Table~\ref{tab:mcdata}).
It might be expected that $\abs{Z\rsub{P}}$ should decrease in response to the 
increasing intermolecular electrostatic repulsion at higher protein 
concentration. 
But the opposite observed trend is due to the more effective screening of 
intramolecular electrostatic repulsion at higher protein concentration (the 
Debye screening length, $\kappa^{-1}$, is controlled by the counterions). 

%
%
\subsection{\label{subsec:colloidsq}Structure Factor from Colloidal Models}

The preceding analysis with the CGSB interaction model indicates that the 
structure factor is governed mainly by excluded volume and screened electrostatic 
interactions. 
To assess the importance of the irregular shape and the inhomogeneous charge 
distribution of the proteins, we consider two colloidal models where the protein 
is described as a sphere. 
These models are conceptually simple and computationally convenient since 
$S\nbr{q}$ can be expressed in analytic form (see Secs.\ \ref{sec:hs} 
and~\ref{sec:yukawa} in Supporting Information). 

The first model is the hard-sphere fluid, where the only interaction is the 
hard-core repulsion and the diameter, $\sigma\rsub{P}$, of the spherical protein 
is fixed by the requirement that the sphere has the same volume as the real 
protein (see Sec.~\ref{subsec:colloid}). 
For IFABP the structure factor predicted by the hard-sphere model is virtually 
identical to that obtained with the CGSB model in the $q$ range accessed by the 
MC simulations (Fig.~\ref{fig:sqpy}c). 
For Mb the agreement between the two models is also good, although the 
hard-sphere $S\nbr{q}$ is slightly displaced to larger $q$ 
(Fig.~\ref{fig:sqpy}b). 
For BPTI, on the other hand, the predictions of the two models differ markedly 
(Fig.~\ref{fig:sqpy}a). 

For Mb and IFABP, the agreement between the two models indicates that shape 
asymmetry and charge inhomogeneity are unimportant under the examined solution 
conditions. 
All three proteins have similar (spheroid) aspect ratios of 1.5--1.6, but 
neither this asymmetry nor the (coarse-grained) surface roughness appears to 
influence $S\nbr{q}$ significantly. 
In contrast to this finding, model calculations of the osmotic second virial 
coefficient, $B_{22}$, for several proteins indicate that while coarse-graining 
at the amino acid level (as in our CGSB model) has little effect (compared to an 
all-atom description), a hard-sphere model (with the same volume as the real 
protein) underestimates $B_{22}$ by $\sim 35\%$.\cite{Grunberger2013} 

The excellent agreement between the two models for IFABP can be further 
rationalized by the nearly zero net charge at the examined pH 
(Table~\ref{tab:mcdata}). 
Thus, at least for this protein, the inhomogeneous distribution of discrete 
charges appears to be unimportant. 
Mb has a larger, but still small, net charge (Table~\ref{tab:mcdata}), which may 
account for slight shift of $S\nbr{q}$ to smaller $q$ values (corresponding to 
longer distances) when the longer-ranged electrostatic repulsion is accounted 
for (in the CGSB model). 
For BPTI at pH~4, where $Z\rsub{P} \approx +7$, electrostatic repulsion 
suppresses $S\nbr{q}$ more than for hard-core repulsion alone and also shifts 
the onset of this suppression to smaller $q$ values, as expected from the longer 
range of the electrostatic repulsion (Fig.~\ref{fig:sqpy}a). 

In a recent SAXS study of BPTI and Mb solutions, Goldenberg and Argyle found 
that the experimental structure factor for Mb (at pH~7) can be well described by 
a hard-sphere model.\cite{Goldenberg2014} 
While this conforms with our findings, it should be noted that these authors 
fitted both the hard-sphere diameter, $\sigma\rsub{P}$, and the protein volume 
fraction, $\phi\rsub{P}$, to the SAXS data. 
For Mb, the fit yielded $\sigma\rsub{P} = 3.74$~nm,\cite{Goldenberg2014} 
slightly larger than the experimentally based value of 3.46~nm used here. 
It should also be noted that the solvent used by Goldenberg and Argyle contained 
1~\M\ urea and 50~\mM\ phosphate buffer.\cite{Goldenberg2014} 
Also for BPTI (at pH~7 with $Z\rsub{P} \approx +6$), the hard-sphere model gave 
reasonable fits to the SAXS data, presumably because the buffer screened out 
most of the electrostatic interactions.\cite{Goldenberg2014} 
But the fitted hard-sphere diameter, $\sigma\rsub{P}$, was found to depend 
strongly on the buffer type, indicating that specific ion binding affects the 
protein--protein interaction.\cite{Goldenberg2014} 

While we cannot compare the two models below $q = 0.5~\inm$ since the MC 
simulations do not access this range, we can compare the hard-sphere model with 
the experimental structure factor. For Mb the experimental $S\nbr{q}$ is 
slightly smaller than for hard spheres (Fig.~\ref{fig:sqpy}b), consistent with 
a modest contribution from electrostatic repulsion. 
The more pronounced discrepancy seen for IFABP (Fig.~\ref{fig:sqpy}c) can 
hardly be attributed to electrostatic repulsion since IFABP has a smaller net 
charge than Mb. 
Possibly, the drop of $S\nbr{q}$ below $q = 0.5~\inm$ is an artifact of 
incorporating the effect of IFABP aggregation in the AFF (\textit{vide supra}). 
 
For the more highly charged protein BPTI, the $S\nbr{q}$ predicted by the 
hard-sphere model differs substantially from the experimental and CGSB-based 
structure factors (Fig.~\ref{fig:sqpy}a). 
We therefore investigated another colloidal interaction model, the HSY fluid, 
with a screened Coulomb repulsion in addition to the hard-core repulsion. 
The HSY model thus includes the two dominant interactions in the CGSB model, 
but the protein is now described as a sphere with a uniform surface charge 
density. 
As for the other models, we do not optimize the model parameters: the net 
charge, $Z\rsub{P} \approx +7$, and the Debye screening length, $\kappa^{-1}$, 
are taken from Table~\ref{tab:mcdata} and the diameter, 
$\sigma\rsub{P} = 2.46$~nm, is fixed by the protein volume (see 
Sec.~\ref{subsec:colloid}), as in the hard-sphere model. 
The structure factor for the HSY model is computed from the analytic MPB-RMSA 
integral equation approximation, which should be quantitatively accurate under 
our conditions.\cite{Heinen2011,Heinen2011b} 

\begin{figure}[t!]
  \centering
  \includegraphics[viewport=0 0 207 286]{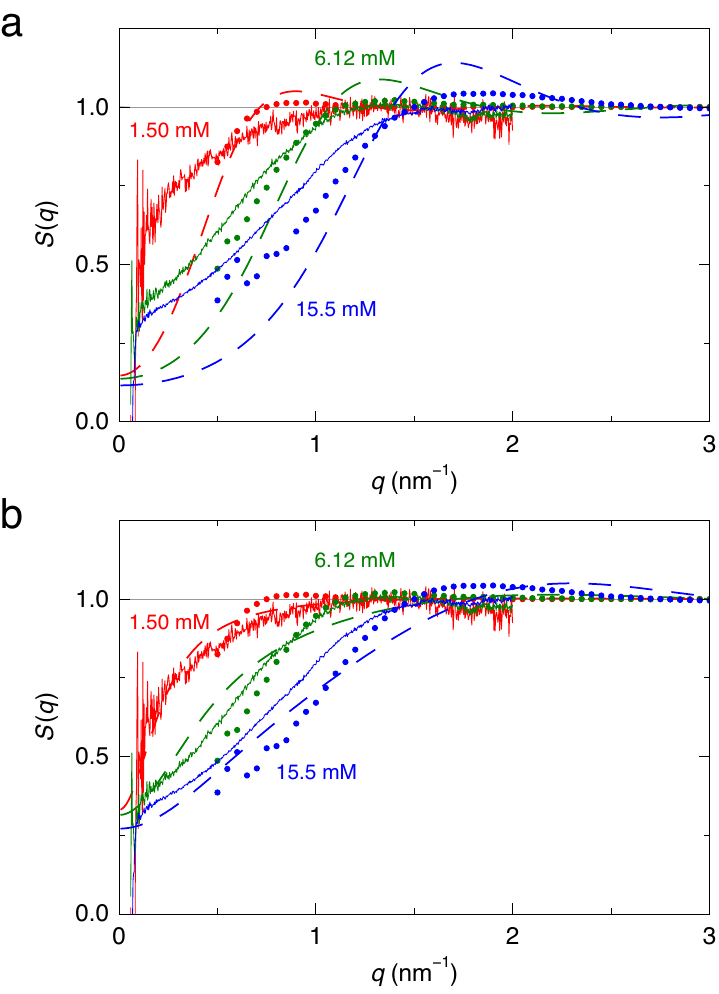}
  \caption{\label{fig:bpti_rmsa}Structure factor for BPTI at three 
           concentrations, obtained from SAXS experiments (solid curves), from 
           the CGSB model without vdW attraction (dots), and from the HSY 
           model (dashed curves). 
           For the latter model, the net charge, $Z\rsub{P}$, was taken from 
           Table~\ref{tab:mcdata} (\textbf{a}) or set to $+2$ (\textbf{b}). 
           The experimental $S\nbr{q}$ is only shown up to $q = 2~\inm$; 
           at higher $q$ the noise amplitude exceeds any deviation from 
           $S\nbr{q} = 1$.} 
\end{figure}

As seen from Fig.~\ref{fig:bpti_rmsa}a, the HSY model produces a too highly 
structured $S\nbr{q}$. 
In other words, the electrostatic repulsion is too strong. 
The agreement with the experimental $S\nbr{q}$ can be improved by reducing the 
net charge (Fig.~\ref{fig:bpti_rmsa}b), but this \textit{ad hoc} modification 
is difficult to justify. 
Since the MPB-RMSA approximation should be accurate, we conclude that the HSY 
model is responsible for the discrepancy. 
Specifically, we infer that the inhomogeneous charge distribution of the real 
protein produces a weaker (orientationally averaged) electrostatic repulsion than the 
same net charge distributed uniformly on a spherical surface. 
Indeed, the crystal structure of BPTI reveals a pronounced charge asymmetry, 
with all the negatively charged carboxylate groups confined to one half of the 
molecule (Fig.~\ref{fig:struct}a). 
For the real protein, the electrostatic interaction should therefore be attractive for 
certain relative orientations so that the effective orientationally averaged 
potential of mean force, $w\nbr{r}$, becomes less 
repulsive.\cite{Striolo2002} 
This anisotropy of the screened electrostatic interaction should also amplify the 
effect on $S\nbr{q}$ of shape asymmetry by favoring close approach of two 
protein molecules for relative orientations with favorable electrostatic interaction. 
This coupling of excluded volume and electrostatic interactions in the potential of 
mean force, $w\nbr{r}$, may be responsible for the observed shift of $S\nbr{q}$ 
to smaller $q$ (larger separations) and the suppressed peak in $S\nbr{q}$, 
relative to the HSY structure factor (Fig.~\ref{fig:bpti_rmsa}). 
Such effects should be less pronounced for Mb and IFABP not only because they 
have smaller net charge, but also because the discrete charge distribution is 
less asymmetric than for BPTI (Fig.~\ref{fig:struct}). 
The HSY structure factors for Mb and IFABP indeed show good agreement with the 
experimental and CGSB $S\nbr{q}$, to the same extent as the hard-sphere model 
(Fig.~\ref{fig:sqpy}), at high $q$ ($\gtrsim 0.5~\inm$) where the coupling 
effect is expected to play an important role 
(Fig.~\ref{fig:mi_rmsa} in Supporting Information). 
Not surprisingly, the charge in the HSY model leads to highly repulsive 
interactions, as in the case of BPTI (Fig.~\ref{fig:bpti_rmsa}a), and the 
model diverges from the experiment at lower $q$ for moderately charged Mb 
(Fig.~\ref{fig:mi_rmsa}). 

\begin{figure}[t!]
  \centering
  \includegraphics[viewport=0 0 199 139]{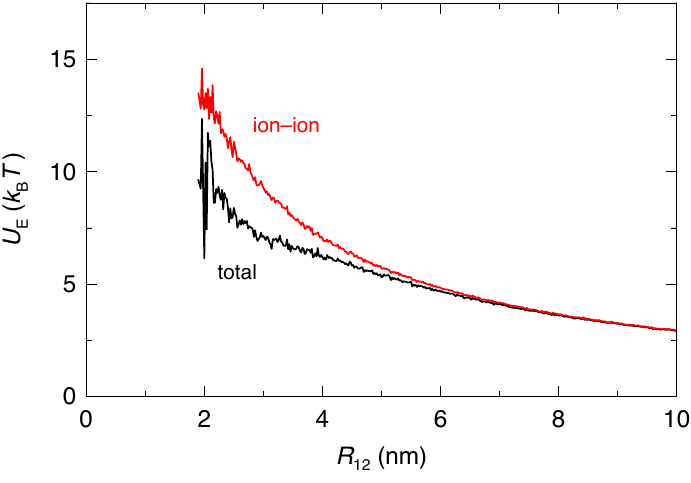}
  \caption{\label{fig:bpti_es}Orientation-averaged electrostatic energy, 
           $U\rsub{E}$, as a function of mass center separation, $R_{12}$, between 
           two BPTI molecules evaluated exactly as 
           $\ensav{\lambda\rsub{B} \sum_i \sum_j z_i z_j/r_{ij}}$ 
           (black), where residues $i$ and $j$ belong to different molecules, 
           and by treating the two proteins as monopoles, 
           $\ensav{\lambda\rsub{B} \sum_i z_i \sum_j z_j / R_{12}}$ 
           (red). 
           The averaging was based on configurations from a two-body MC 
           simulation at pH~4.1 and a Debye length, $\kappa^{-1}$, of 4.37~nm 
           (\textit{cf.}\ Table~\ref{tab:mcdata}).}
\end{figure}

To examine the effect of charge and shape asymmetry on the electrostatic contribution 
to the potential of mean force, 
we performed CGSB MC simulations with only two BPTI molecules at fixed mass-center 
separation and at constant pH. 
From the sampled orientational configurations, we calculated the 
orientation-averaged total (residue-based) electrostatic interaction 
energy between the two molecules and the intermolecular ion--ion interaction 
energy (Fig.~\ref{fig:bpti_es}). 
Note that the CGSB model incorporates both charge and shape asymmetry. 
As seen from Fig.~\ref{fig:bpti_es}, the total electrostatic repulstion is 
weaker than the ion--ion repulsion at short intermolecular separations,  
where charge and shape asymmetry is expected to be important 
(\textit{vide supra}). 

%
%
\section{\label{sec:conc}Conclusions}

From SAXS experiments at multiple protein concentrations, we have determined the 
structure factor for the three globular proteins BPTI, Mb, and IFABP. 
Information about the protein--protein potential of mean force, averaged over 
relative protein orientations and solvent configurations, was derived from the 
experimental structure factors with the aid of several interaction models. 
For a structure-based interaction model coarse-grained to the amino-acid residue 
level, we computed the structure factor by MC simulation. 
For the hard-sphere and HSY models, the structure factor was obtained from 
accurate integral equation approximations. 
The parameters in these interaction models were fixed by the known properties of 
the protein solutions, rather than being optimized for agreement with the SAXS 
data.

For these proteins and under the investigated solution conditions, we find that 
the structure factor can be accounted for by excluded volume and screened 
electrostatic interactions, with no need to invoke other short-ranged, soft 
interactions, such as vdW attraction, hydrophobic and other solvent-related 
interactions. 
We cannot exclude the possibility that the effects on the structure factor of 
some of these apparently unimportant interactions tend to cancel out.     

For Mb and IFABP, with small net charge, the structure factor is well described 
by a hard-sphere model, even though these proteins are non-spherical (aspect 
ratio 1.5--1.6) and contain many charged residues. 
For BPTI, with larger net charge, screened electrostatic repulsion is important, 
but it is weaker than predicted by a HSY model. 
The reduction of the electrostatic repulsion may be a result of the pronounced 
asymmetry of the surface charge distribution for this protein, which tends to 
favor protein--protein encounters with less repulsive electrostatic 
interactions. 

The MC simulations were performed at constant pH and therefore allow for thermal 
fluctuations in the protonation state of ionizable residues. 
Such charge fluctuations do not, however, have a significant effect on the 
protein--protein potential of mean force under the conditions investigated here. 
  
%
%
\section*{Acknowledgments}

{\small
We thank Hanna Nilsson and Annika Rogstam (Lund Protein Production Platform) for 
protein preparation and purification, Marco Heinen for helpful correspondence, 
Bayer Healthcare AG for a generous supply of BPTI, MAX-lab for beamtime on the 
SAXS beamline I911-4 (proposal ID 20120020), LUNARC for computational time, and 
the Swedish Research Council, the Swedish Foundation for Strategic Research, 
Organizing Molecular Matter, eSSENCE in Lund, and the Wenner-Gren Foundations for 
financial support. 
}

%
%

%
%
\onecolumn
\section*{Supporting Information}

\setcounter{section}{0}
\setcounter{figure}{0}
\setcounter{table}{0}
\setcounter{equation}{0}

\renewcommand{\theHsection}{sup.\the\value{section}}
\renewcommand{\thesection}{S\arabic{section}}
\renewcommand{\thefigure}{S\arabic{figure}}
\renewcommand{\thetable}{S\arabic{table}}
\renewcommand{\theequation}{S\arabic{equation}}

%
%
\section{\label{sec:hs}Hard-Sphere Fluid}

For a fluid of identical hard spheres of diameter $\sigma$, the pair interaction 
energy is
\begin{equation}
  w\nbr*{x} = \begin{dcases}
                \infty &,\ x < 1 \ , \\
                0      &,\ x > 1 \ ,
               \end{dcases}
  \label{eqS1}
\end{equation}
where $x = r/\sigma$ is the reduced inter-particle separation.
For this model, the pair correlation function (PCF), $g\nbr{x}$, obeys the exact 
condition
\begin{equation}
  g\nbr{x} = 0 \ ,\ x < 1 \ ,
  \label{eqS2}
\end{equation}
which simply expresses the impenetrability of the hard spheres. 

According to the Percus--Yevick (PY) approximation,\cite{Percus1958} the direct 
correlation function, $c\nbr{x}$, is related to the PCF and the pair potential 
as
\begin{equation}
  c\nbr*{x} = g\nbr*{x}\cbr*{1-\exp\sbr*{\frac{w\nbr*{x}}{k\rsub{B}T}}} \ .
  \label{eqS3}
\end{equation}
For the hard-sphere model in Eq.~\eqref{eqS1}, this implies that
\begin{equation}
  c\nbr*{x} = \begin{dcases}
                -y\nbr*{x} &,\ x < 1 \ , \\
                0          &,\ x > 1 \ ,
              \end{dcases}
  \label{eqS4}
\end{equation}
where the function 
$y\nbr{x} \equiv g\nbr{x} \exp\sbr{w\nbr{x}/\nbr{k\rsub{B}T}}$ is continuous at 
$x = 1$.

For the hard-sphere fluid, the approximate PY closure in Eq.~\eqref{eqS4} allows 
the formally exact Ornstein--Zernike (OZ) integral equation\cite{Hansen1986} to 
be solved analytically.\cite{Wertheim1964}
The resulting structure factor\cite{Nagele2004} is a function of the reduced 
wavevector $Q \equiv q\sigma$ and the particle volume fraction 
$\phi = n\rsub{P}\pi\sigma^3/6$:
\begin{equation}
  S\nbr*{Q} = \frac{1}{\abs*{F\nbr*{Q}}^2} 
            = \frac{1}{\sbr*{\mathrm{Re}\,F\nbr*{Q}}^2 
                       + \sbr*{\mathrm{Im}\,F\nbr*{Q}}^2} \ ,
  \label{eqS5}
\end{equation}
with
\begin{equation}
  {\allowdisplaybreaks\begin{split}
    \mathrm{Re}\,F\nbr*{Q} &= 
      1 - 12\phi\sbr*{a_0\nbr*{\phi} G_a\nbr*{Q} 
                      + b_0\nbr*{\phi} G_b\nbr*{Q}} \ , \\
    \mathrm{Im}\,F(Q) &= 
      -12\phi\sbr*{a_0\nbr*{\phi} H_a\nbr*{Q} 
                   + b_0\nbr*{\phi} H_b\nbr*{Q}}    \ .
  \end{split}}
  \label{eqS6}
\end{equation}
Here we have defined
\begin{equation}
  {\allowdisplaybreaks\begin{split}
    a_0\nbr*{\phi} &= \frac{1 + 2\phi}{\nbr*{1-\phi}^2} \ , \\
    b_0\nbr*{\phi} &= -\frac{3\phi}{2\nbr*{1-\phi}^2}   \ ,
  \end{split}}
  \label{eqS7}
\end{equation}
and
\begin{equation}
  {\allowdisplaybreaks\begin{split}
    G_a\nbr*{Q} &= \frac{Q\cos Q - \sin Q}{Q^3}             \ , \\
    G_b\nbr*{Q} &= \frac{\cos Q - 1}{Q^2}                   \ , \\
    H_a\nbr*{Q} &= \frac{Q\sin Q + \cos Q - 1 - Q^2/2}{Q^3} \ , \\
    H_b\nbr*{Q} &= \frac{\sin Q - Q}{Q^2} \ .
  \end{split}}
  \label{eqS8}
\end{equation}
This analytic result is highly accurate up to volume fractions 
$\phi \approx 0.35$.

%
%
\section{\label{sec:yukawa}Hard-Sphere Yukawa Fluid}

Solutions of charged colloidal particles or proteins are often modeled as a 
one-component macrofluid composed of charged hard spheres in a uniform 
neutralizing background medium. 
Apart from their excluded volume, the particles are taken to interact with a 
screened Coulomb (Yukawa) potential, so that
\begin{equation}
  \frac{w\nbr*{x}}{k\rsub{B}T} = 
    \begin{dcases}
      \infty                   &,\ x < 1 \ , \\
      \gamma \frac{e^{-kx}}{x} &,\ x > 1 \ ,
    \end{dcases}
  \label{eqS9}
\end{equation}
where $x \equiv r/\sigma$. Furthermore, $\gamma$ is a dimensionless coupling 
constant and $k$ is a dimensionless screening parameter. 
These are given by
\begin{equation}
  \gamma \equiv \frac{\lambda\rsub{B}}{\sigma} 
                \frac{e^k}{\nbr*{1+k/2}^2} Z^2 \ ,
  \label{eqS10}
\end{equation}
\begin{equation}
  k^2 \equiv \nbr*{\kappa\sigma}^2 
    = 4\pi\lambda\rsub{B}\sigma^2 
        \nbr*{n\rsub{P}\abs*{Z} + 2n\rsub{S}}
    = \frac{\lambda\rsub{B}}{\sigma} 
        \nbr*{24\phi\abs*{Z} + 8\pi n\rsub{S} \sigma^3} \ ,
  \label{eqS11}
\end{equation}
where $Z$ is the net protein charge (in units of $e$), $n\rsub{P}$ is the 
protein number density, $\phi = n\rsub{P} \pi \sigma^3/6$ is the protein volume 
fraction, and $n\rsub{S}$ is the number density of monovalent salt. 
The number density of counterions, also assumed monovalent, is 
$n\rsub{P} \abs{Z}$.
Finally, 
$\lambda\rsub{B} = e^2/\nbr{4 \pi \e_0 \e\rsub{r} k\rsub{B} T}$ 
is the Bjerrum length.

For the hard-sphere Yukawa (HSY) model in Eq.~\eqref{eqS9}, the OZ equation can 
be solved analytically in the mean spherical approximation 
(MSA),\cite{Waisman1973,Hoye1977} defined by 
\begin{equation}
    c\nbr*{x} = -\frac{w\nbr*{x}}{k\rsub{B}T} \ ,\ x > 1 \ ,
    \label{eqS12}
\end{equation}
along with Eq.~\eqref{eqS2}.
Various forms of the lengthy analytic solution for the structure factor, 
$S\nbr{q}$, and other quantities have been published.
All involve the solution of a quartic equation, but different formulations have 
been presented where the quartic equation takes different forms. 
The results of Hayter and Penfold\cite{Hayter1981} are free from misprints, but 
they do not provide an analytic solution of the quartic equation. 
Moreover, the physical root (among the four possible roots) is identified by 
showing that it leads to $g\nbr{x} = 0$ for $x < 1$ and this requires a 
numerical Fourier transform. 
For analysis of SAXS data, it is more convenient to use the completely analytic 
formulation presented by Cummings 
\textit{et al}.\cite{Cummings1979a,Cummings1979b,Cummings1983,Heinen2011b}   

In this so-called Wiener--Hopf factorization approach, a complex-valued function 
$F\nbr{Q}$ is defined such that 
\begin{equation}
  1 - n\rsub{P} \hat{c}\nbr*{Q} = F\nbr*{Q} F\nbr*{-Q} 
    = \abs*{F\nbr*{Q}}^2 
    = \sbr*{\mathrm{Re}\,F\nbr*{Q}}^2 
            + \sbr*{\mathrm{Im}\,F\nbr*{Q}}^2 \ ,
  \label{eqS13}
\end{equation}
where $Q \equiv q \sigma$ and $F\nbr*{-Q} = \sbr{F\nbr{Q}}^\ast$ for real $Q$. 
The structure factor, $S\nbr{Q}$, can then be expressed on the form of 
Eq.~\eqref{eqS5}.
The function $F\nbr{Q}$ is related to the Fourier transform of another function 
$F\nbr{x}$:
\begin{equation}
  F\nbr*{Q} = 1 - 2 \pi n\rsub{P} \sigma^3 
                  \int_{-\infty}^{\infty}\dd x\,e^{iQx} F\nbr*{x}
            = 1 - 12\phi 
                  \int_{-\infty}^{\infty}\dd x\,e^{iQx} F\nbr*{x} \ .
  \label{eqS14}
\end{equation}
For the HSY model in the MSA approximation, the function $F\nbr{x}$ is given 
by\cite{Cummings1983} (but Cummings' earlier 
papers\cite{Cummings1979a,Cummings1979b} give this function incorrectly) 
\begin{equation}
  F\nbr*{x} =
    \begin{dcases}
      0                                    &,\ x < 0       \ , \\
      F_0\nbr*{x} + \beta e^{-k\nbr*{x-1}} &,\ 0 \le x < 1 \ , \\
      \beta e^{-k\nbr*{x-1}}               &,\ x \ge 1     \ ,
    \end{dcases}
  \label{eqS15}
\end{equation}
where $k$ is defined by Eq.~\eqref{eqS11} and
\begin{equation}
  F_0\nbr*{x} = \frac{a}{2}\nbr*{x^2-1}+b\nbr*{x-1} 
                    + d\beta\sbr*{1 - e^{-k\nbr*{x-1}}} \ .
  \label{eqS16}
\end{equation}
The quantities $a$, $b$, $d$, and $\beta$ are functions of the system parameters 
$\gamma$, $k$, and $\phi$.

Combining Eqs.\ \eqref{eqS14}--\eqref{eqS16} and performing the integral, one 
obtains
\begin{equation}
  {\allowdisplaybreaks\begin{split}
    \mathrm{Re}\,F\nbr*{q} &= 1 - 12\phi 
      \sbr*{a G_a\nbr*{Q} + b G_b\nbr*{Q} 
            + \frac{\beta U\nbr*{Q}}{\nbr*{k^2 + Q^2}}} \ , \\
     \mathrm{Im}\,F\nbr*{q} &= -12\phi 
       \sbr*{a H_a\nbr*{Q} + b H_b\nbr*{Q} 
             + \frac{\beta V\nbr*{Q}}{\nbr*{k^2 + Q^2}}} \ ,
  \end{split}}
  \label{eqS17}
\end{equation}
where the functions $G_a\nbr{Q}$, $G_b\nbr{Q}$, $H_a\nbr{Q}$, and $H_b\nbr{Q}$ 
are given by Eq.~\eqref{eqS8} and
\begin{equation}
  {\allowdisplaybreaks\begin{split}
    U\nbr*{Q} &= d k \sbr*{\cos Q + k \frac{\sin Q}{Q}} 
                   - \nbr*{d-1} k e^k            \ , \\
    V\nbr*{Q} &= d k \sbr*{\sin Q - k \frac{\nbr*{\cos Q - 1}}{Q}} 
                   + Q \sbr*{d - \nbr*{d-1} e^k} \ .
  \end{split}}
  \label{eqS18}
\end{equation}
The quantities $a$ and $b$ are given by\cite{Cummings1979a,Cummings1979b,Heinen2011b}
\begin{equation}
  a = a_0 + \frac{12 \phi \beta}{k \nbr*{1-\phi}^2} 
              \sbr*{3 d \phi k - \omega \tau} \ ,
  \label{eqS19}
\end{equation}
\begin{equation}
  b = b_0 + \frac{12 \phi \beta}{k \nbr*{1-\phi}^2} 
              \sbr*{\nbr*{1-4\phi} \frac{kd}{2} + \omega\rho} \ ,
  \label{eqS20}
\end{equation}
with $a_0$ and $b_0$ as defined in Eq.~\eqref{eqS7}, and
\begin{equation}
  \omega = d \nbr*{1+k} - \nbr*{d-1} e^k \ ,
  \label{eqS21}
\end{equation}
\begin{equation}
  \tau = 1 + 2\phi - \frac{6\phi}{k} \ ,
  \label{eqS22}
\end{equation}
\begin{equation}
  \rho = \frac{3\phi}{2} + \frac{\nbr*{1-4\phi}}{k} \ .
  \label{eqS23}
\end{equation}

The dependence on the coupling constant $\gamma$ enters via the quantity
\begin{equation}
  d = \frac{\nbr*{\beta D+K} e^{-k} + \beta^2 E}{\beta^2 F} \ ,
  \label{eqS24}
\end{equation}
which involves the additional quantities
\begin{equation}
  D = k - a_0 P - b_0 T \ ,
  \label{eqS25}
\end{equation}
\begin{equation}
  E = -6 \phi + \delta \nbr*{\tau P - \rho T} \ ,
  \label{eqS26}
\end{equation}
\begin{equation}
  F = -6 \phi \nbr*{1 - e^{-k}}^2 
        + P \delta \sbr*{\mu \tau + 3\phi k e^{-k}}
        - T \delta \sbr*{\mu \rho - \frac{1}{2} \nbr*{1-4 \phi} k e^{-k}} \ ,
  \label{eqS27}
\end{equation}
where
\begin{equation}
  \delta = \frac{12 \phi}{k \nbr*{1-\phi}^2} \ ,
  \label{eqS28}
\end{equation}
\begin{equation}
  \mu = 1 - \nbr*{1+k} e^{-k} \ ,
  \label{eqS29}
\end{equation}
\begin{equation}
  P = 12 \phi \nbr*{\frac{\mu}{k^2} - \frac{1}{2}} \ ,
  \label{eqS30}
\end{equation}
\begin{equation}
  T = \frac{12 \phi}{k} \nbr*{1 - k - e^{-k}} \ ,
  \label{eqS31}
\end{equation}
and the (non-negative) coupling strength parameter
\begin{equation}
  K = \gamma e^{-k} \ .
  \label{eqS32}
\end{equation}

Finally, $\beta$ is one of the four roots of the quartic equation
\begin{equation}
  36 \phi^2 \beta^4 - 6 \phi X \beta^3 - 12 \phi K \beta^2 
    + K Y \beta + K^2 = 0 \ ,
  \label{eqS33}
\end{equation}
where
\begin{equation}
  {\allowdisplaybreaks\begin{split}
    X &= k e^{-k} - \frac{6 \phi}{\nbr*{1-\phi} k^2} 
           \sbr*{2 - 2 k - \nbr*{2-k^2} e^{-k}} - W \ , \\
    Y &= k - \frac{6 \phi}{\nbr*{1-\phi} k^2} 
           \sbr*{2 - k^2 - 2 \nbr*{1+k} e^{-k}} - W \ , \\
    W &= \frac{18 \phi^2}{\nbr*{1-\phi}^2 k^2} 
           \sbr*{2 - k - \nbr*{2+k} e^{-k}} \ .
  \end{split}}
  \label{eqS34}
\end{equation}
The desired root reduces to the PY solution in the limit $K \rightarrow 0$ and 
in the limit $\phi \rightarrow 0$ it yields\cite{Cummings1979a,Cummings1979b}
\begin{equation}
  \beta = -\frac{K}{k} \sbr*{1 + \mathcal{O}\nbr*{\phi}} \ .
  \label{eqS35}
\end{equation}
We obtain the following analytic expression for the desired root:
\begin{equation}
  {\allowdisplaybreaks\begin{split}
    \beta =\;& \frac{X}{24 \phi} 
               + \frac{\sqrt{B_2} - \sqrt{B_1}}{2} \ , \\
    B_1 =\;& \Gamma + \Lambda \ , \\
    B_2 =\;& 2 \Gamma - \Lambda 
             - \frac{1}{18 \phi^2 \sqrt{B_1}} 
                 \sbr*{K \nbr*{X-Y} + \frac{X^3}{48 \phi}} \ , \\
    \Gamma =\;& \sbr*{\frac{X}{12 \phi}}^2 + \frac{2 K}{9 \phi} \ , \\
    \Lambda =\;& \frac{R}{\nbr*{6 \phi}^2} + C \ , \\
    C =\;& \frac{K \nbr*{32 K \phi + X Y}}{18 \phi R} \ , \\
    R =\;& \cbr*{2 K \phi^2 \sbr*{M - \nbr*{\frac{K L}{\phi}}^{\!\!1/2}}}^{\!\!1/3} \ , \\
    M =\;& \nbr*{16 K}^2 \phi + 3 K \nbr*{3 X^2 + 3 Y^2 - 2 X Y} \ , \\
    L =\;& 2 \sbr*{48 K \phi \nbr*{X-Y}}^2
           - 2 \nbr*{X Y}^3 \\
         & + 3 K \phi \sbr*{2 \nbr*{X Y}^2 
           + 9 \nbr*{3 X^4 - 4 X^3 Y - 4 X Y^3 + 3 Y^4}} \ .
  \end{split}}
  \label{eqS36}
\end{equation}

Apart from the errors\cite{Cummings1979a,Cummings1979b} in $F\nbr{x}$ noted 
above, the literature contains several other misprints in the analytic MSA 
result. 
Cummings \& Smith in their Molecular Physics paper\cite{Cummings1979a} have a 
sign error in $W$ (their Eq.~(6a)), defined here in Eq.~\eqref{eqS34}, and in 
their Chemical Physics paper\cite{Cummings1979b} they omit a factor 
$\exp\nbr{k}$ in the definition of $b$ (their Eq.~(9b)). 
Marco Heinen in his PhD thesis\cite{Heinen2011b} (Appendix A) has misprints in 
the second lines of his Eqs.\ (A.2) and~(A.3) (which define 
$\mathrm{Re}\,F\nbr{q}$ and $\mathrm{Im}\,F\nbr{q}$), where $b$ appears in place 
of $k$. 
Heinen also introduces a quantity $f$, which is unnecessary since 
$f = \nbr{1-d} \exp\nbr{k}$. 
Furthermore, all other authors define $K$ with the opposite sign to that in 
Eq.~\eqref{eqS32}.

The MSA solution of the HSY model is accurate (as compared to Monte Carlo 
simulations of the same model) for weakly charged macroions at relatively high 
volume fractions. 
But for highly charged macroions and/or at low volume fractions, the MSA 
produces unphysical results. 
Specifically the contact PCF, $g\nbr{\sigma}$, becomes negative. 
Various schemes have been proposed to improve the MSA. 
The basic idea is that, under the conditions where the MSA fails, the macroions 
are almost always so far apart (because the volume fraction is low and/or 
because of strong electrostatic repulsion) that the actual hard-sphere diameter 
$\sigma$ has no effect on $S\nbr{q}$. 
It is therefore possible to increase $\sigma$ to a larger value $\sigma'$ so 
that $g\nbr{\sigma}$ remains non-negative. 
Specifically, $\sigma'$ is chosen so that $g(\sigma';\phi') = 0$, where 
$\phi' = \phi \nbr{\sigma'/\sigma}^3$ is the rescaled volume fraction. 
(The volume fraction increases because the particle size is increased at 
constant particle number density $n\rsub{P}$.) 
This approach is called the rescaled MSA (RMSA).\cite{Hansen1982}

Comparison with computer simulations shows that even the RMSA is not accurate 
for strongly repulsive macroions (high charge and/or low salt concentration). 
In particular, the RMSA tends to underestimate the local ordering by yielding a 
too small principal peak in $S\nbr{q}$ (and in $g\nbr{r}$) and a too large 
osmotic compressibility, $S\nbr{0}$. 
It was shown that the accuracy of the RMSA can be further improved by redefining 
the model parameters $\gamma$ and $k$ to correct for the fact that the 
counterions are treated in the one-component macrofluid model (of which the HSY 
model is a special case) as a uniform background medium that penetrates the 
macroion and therefore reduces its effective charge. 
This scheme is called the penetrating-background corrected RMSA 
(PB-RMSA).\cite{Snook1992}
A further improvement, yielding a structure factor, $S\nbr{q}$, in excellent 
agreement with Monte Carlo simulations in the full parameter space, was obtained 
with a modified PB-RMSA (MPB-RMSA) scheme.\cite{Heinen2011} 
This MPB-RMSA scheme involves the following steps:\cite{Heinen2011}

\begin{enumerate}[label=(\arabic*),series=conc]
  \item Specify the true model parameters $\sigma$, $\phi$, $\gamma$, and $k$, 
        with $\gamma$ given by Eq.~\eqref{eqS10} and $k$ by the following 
        modified version of Eq.~\ref{eqS11}:
        \begin{equation}
          k^2 = \frac{\lambda\rsub{B}}{\sigma \nbr*{1-\phi}} 
                  \nbr*{24 \phi \abs*{Z} + 8 \pi  n\rsub{S} \sigma^3}\ .
          \label{eqS37}
         \end{equation}

  \item Compute the modified parameters:
        \begin{equation}
          {\allowdisplaybreaks\begin{split}
            k\rsub{mod} &= k \nbr*{1-\phi}^{1/2} \ , \\
            \gamma\rsub{mod} &= \gamma \exp\nbr*{k\rsub{mod}-k} 
                                  \nbr*{\frac{1+k/2}{1+k\rsub{mod}/2}}^{\!\!2} \ .
          \end{split}}
          \label{eqS38}
        \end{equation}

  \item Assign the further modified parameters:
        \begin{equation}
          {\allowdisplaybreaks\begin{split}
            k^\ast &= k\rsub{mod} - 2 \phi^{1/3} \ln\nbr*{1-\phi} \ , \\
            \gamma^\ast &= \frac{\gamma\rsub{mod}}{\nbr*{1-\phi}^2} \ .
          \end{split}}
          \label{eqS39}
        \end{equation}

  \item The contact PCF is given by\cite{Heinen2011b}
        \begin{equation}
          {\allowdisplaybreaks\begin{split}
            g_0 \equiv g\nbr*{\sigma} 
              =& a \nbr*{1 + \beta P} + b \nbr*{1+\beta T} - K 
                 + \beta k \nbr*{d-1} \\
               & - 6 \phi \beta^2 \sbr*{2 d\nbr*{\cosh k - 1} - e^k} \ .
          \end{split}}
          \label{eqS40}
        \end{equation}
        Using this result, compute $g_0\nbr*{\phi,\gamma^\ast,k^\ast}$, that is, 
        the contact PCF with modified Yukawa parameters. 
        If $g_0 < 0$, assign an initial rescaling parameter $s = 0.99$ and go to 
        step~5.
        If $g_0 \ge 0$, set $\sigma^\ast = \sigma$ and $\phi^\ast = \phi$ and go 
        to step~7.

  \item Compute the rescaled parameters:
        \begin{equation}
          {\allowdisplaybreaks\begin{split}
            \sigma^\ast &= \frac{\sigma}{s} \ , \\
            \phi^\ast &= \frac{\phi}{s^3} \ , \\
            \gamma^\ast &= \frac{\gamma\rsub{mod} s}{\nbr*{1-\phi^\ast}^2} \ , \\
            k^\ast &= \frac{k\rsub{mod}}{s} - 2 \nbr*{\phi^\ast}^{1/3} 
                        \ln \nbr*{1-\phi^\ast} \ .
          \end{split}}
          \label{eqS41}
        \end{equation}

  \item Using Eq.~\eqref{eqS40}, compute 
        $g_0\nbr{s} \equiv g_0\nbr{\phi^\ast,\gamma^\ast,k^\ast}$, that is, the 
        contact PCF with rescaling parameter $s$. 
        If $\abs{g_0\nbr{s}} <$ \texttt{tol}, go to step~7. 
        Here \texttt{tol} is a tolerance parameter with default value $10^{-6}$.
        If $\abs{g_0\nbr{s}} \ge$ \texttt{tol}, compute a new rescaling 
        parameter $s'$ that yields a $g_0\nbr{s'}$ closer to $0$. 
        This is accomplished by iteratively solving the equation 
        $g_0\nbr{s'} = 0$ with the Newton--Raphson algorithm:
        \begin{equation}
          s' = s - g_0\nbr*{s} 
                 \sbr*{\frac{\dd g_0\nbr*{s}}{\dd s}}^{-1} \ .
          \label{eqS42}
        \end{equation}
        Then set $s = s'$ and go to step~5.

  \item Compute the structure factor, $S\nbr{q}$, in the desired $q$ range by 
        using Eqs.\ \eqref{eqS5}, \eqref{eqS7}, \eqref{eqS8}, 
        \eqref{eqS17}--\eqref{eqS32},  and \eqref{eqS36} and the input 
        parameters $Q^\ast = q\sigma^\ast$, $\phi^\ast$, $\gamma^\ast$, and 
        $k^\ast$.

\end{enumerate}

\vspace*{\fill}
\begin{figure}[h!]
  \centering
  \includegraphics[viewport=0 0 207 286]{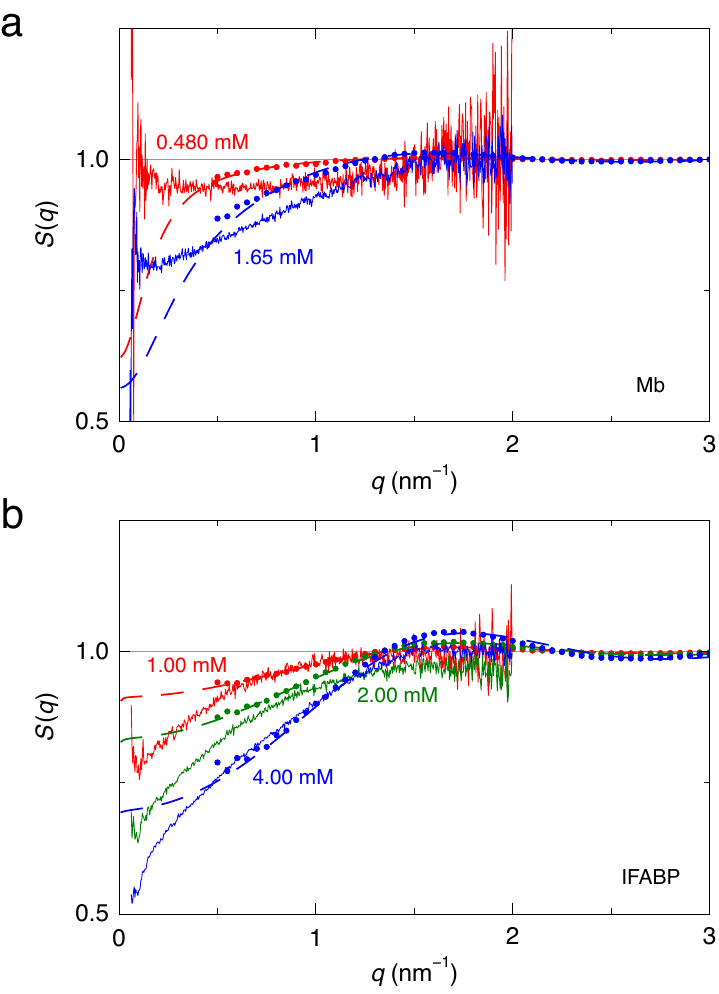}
  \caption{\label{fig:mi_rmsa}Structure factor for Mb (\textbf{a}) and 
           IFABP (\textbf{b}) solutions at multiple concentrations, 
           obtained from SAXS experiments (solid curves), from the CGSB model 
           without vdW attraction (dots), and from the HSY model 
           (dashed curves). 
           For the HSY model, the net charge, $Z\rsub{P}$, was taken from 
           Table~\ref{tab:mcdata}.
           The experimental $S(q)$ is only shown up to $q = 2~\inm$; at higher 
           $q$ the noise amplitude exceeds any deviation from $S\nbr{q} = 1$.} 
\end{figure}
\vspace*{\fill}

\end{document}